\documentclass[aps,preprint,superscriptaddress,groupedaddress]{revtex4}  
\usepackage{graphicx}  
\usepackage{dcolumn}   
\usepackage{bm}        
\usepackage{amssymb}   

\begin{document}



\title{Efficiency of cellular uptake of nanoparticles via receptor-mediated endocytosis}

\author{Anand Banerjee}
\affiliation{Program in Physical Biology, Eunice Kennedy Shriver National Institute of Child Health, and Human Development, National Institutes of Health, Bethesda, Maryland 20892, USA}

\author{Alexander Berzhkovskii}
\affiliation{Mathematical and Statistical Computing Laboratory, Division of Computational Bioscience, Center for Information Technology, National Institutes of Health, Bethesda, Maryland 20892, USA}

\author{Ralph Nossal}
\affiliation{Program in Physical Biology, Eunice Kennedy Shriver National Institute of Child Health, and Human Development, National Institutes of Health, Bethesda, Maryland 20892, USA}

\date{\today}

\begin{abstract}

Experiments show that cellular uptake of nanoparticles, via receptor-mediated 
  endocytosis, strongly depends on nanoparticle size. There is 
  an optimal size, approximately 50\,nm in diameter, at which
  cellular uptake is the highest.  In addition, there is a maximum size,
  approximately 200\,nm, beyond which uptake via receptor-mediated 
  endocytosis does not occur. By comparing results from different 
  experiments, we found that these sizes weakly depend on the type of cells,
  nanoparticles, and ligands used in the experiments. Here, 
  we argue that these observations are consequences of the 
  energetics and assembly dynamics of the protein coat that forms on
  the cytoplasmic side of the outer cell membrane during receptor-mediated 
  endocytosis. Specifically, we show that the
 energetics of coat formation imposes an upper bound on the size of the nanoparticles
 that can be internalized, whereas the nanoparticle-size-dependent dynamics of coat 
 assembly results in the optimal nanoparticle size. The weak dependence of the
 optimal and maximum sizes on cell-nanoparticle-ligand type also follows naturally 
  from our analysis.

\vspace{0.5in}

\leftline{email - banerjeea3@mail.nih.gov}

\end{abstract}
\maketitle

\section*{INTRODUCTION}
In recent years there has been great interest in using nanoparticles
(NPs) for various biomedical applications including imaging,
biosensing, and targeted gene/drug delivery (see review articles
\cite{Sahoo2003,Bareford2007,Akinc2013} and references therein).  
Successful realization of these applications requires efficient cellular uptake
of the NPs. To this end, the NPs are coated with ligands that allow
them to bind to specific cell surface receptors and be internalized
via receptor-mediated endocytosis. An understanding of how the
physical properties of NPs, like their size, shape, charge, etc.,
affect the internalization process is crucial for designing NPs for
biomedical purposes.

Experiments show that the NP size is an important parameter that
determines the mechanism of their cellular uptake.  In particular, NPs
smaller than approximately 200\,nm are internalized typically via
receptor-mediated endocytosis whereas, for larger NPs other
mechanisms are involved \cite{Rejman2004d,Oh2009}.  Furthermore, the
uptake rate of the NPs, that are internalized via receptor-mediated endocytosis, is strongly
size dependent. There is an optimal NP size, approximately 50\,nm in
diameter, at which the uptake rate is highest
\cite{Osaki2004,Rejman2004d,Chithrani2006,Wang2010,Lu2009a,Oh2009,Albanese2011,Andar2013}. In
Table\,\ref{tab1} we collected data from various experiments designed
to study size dependence of NP uptake. The experiments were performed
using different kinds of cell lines, NPs, and ligands; yet the optimal
size was found to be approximately the same. This surprising
observation suggests that the optimal size weakly depends on the above
mentioned factors.

Several theoretical models of receptor-mediated endocytosis of NPs have been proposed in the
literature \cite{Gao2005,Zhang2009,Yuan2010,Decuzzi2007,Mirigian2013}.
A common feature of these models is that they assume the uptake is
controlled by the formation of chemical bonds between receptors
on the cell surface and ligands attached to the NP. Using such an
approach these studies conclude that the optimal NP size is a function
of the receptor density on the cell membrane, ligand density on the
NP, and the receptor-ligand binding energy. These parameters however
can change significantly depending on the cell line and ligands used
in the experiment.  Therefore, in the framework of these models, it is
difficult to explain the same optimal size observed in different
experiments.  Furthermore, such an approach leads to the prediction
that even micron sized NPs can be internalized via receptor-mediated endocytosis
\cite{Gao2005,Zhang2009} which has never been observed
experimentally.

Along with the formation of chemical bonds between the NP ligands and
cell surface receptors, receptor-mediated endocytosis involves the assembly 
of a protein coat on
the cytoplasmic side of the outer cell membrane. In the case of
clathrin-mediated endocytosis - which is a form of receptor-mediated 
endocytosis - the coat
contains several proteins including clathrin, adaptor proteins,
membrane bending proteins like epsin, amphiphysin,
etc.\cite{Mousavi2004,Ungewickell2007}.  The coat assembly plays 
a vital role in internalization of cargo, and any interference with this
process drastically reduces the endocytic capacity of a cell. For
example, cellular uptake of NPs is significantly reduced when the
cells are pretreated with sucrose or potassium-depleted medium
\cite{Rejman2004d,Chithrani2007}, which are known to disrupt coat
formation.  The above mentioned models do not take the coat assembly
into explicit consideration. Therefore, how this key aspect of the
cellular endocytic machinery affects the uptake of NPs remains to be
elucidated.

\begin{table}[b]
\begin{center}
\caption{Summary of experimental results on the size dependence of
  cellular uptake of NPs.  NPs smaller than 200\,nm are internalized
  via receptor-mediated endocytosis, whereas for larger NPs other
  internalization mechanisms are involved
  \cite{Rejman2004d,Oh2009}. Bold faced numbers indicate the NP size
  for which the cellular uptake is highest.  }\label{tab1}
\begin{tabular}{ccccc}\hline
Cells                           &         NP type   &    Ligand   & NP size (diameter nm)      &     Ref. 
\\ \hline
B16-F10                    &   Latex beads &  No ligand   & {\bf 50},100,200,500  &  \citenum{Rejman2004d} \\   
MNNG/HOS            &    Metal hydroxide & Not specified   & {\bf 50},100,200,350  & \citenum{Oh2009} \\    
Hela                           &   Quantum dots &  Not specified   & 5,15,{\bf 50}     &  \citenum{Osaki2004} \\ 
Hela, STO, SNB19    &   Gold  &  Transferrin   & 14,30,{\bf 50},74,100   &  \citenum{Chithrani2006} \\ 
CL1-0, Hela               &   Gold  & single-stranded DNA   & {\bf 45},75,110  & \citenum{Wang2010} \\  
HeLa                          &   Mesoporous silica & Not specified   & 30,{\bf 50},110,170,280   & \citenum{Lu2009a} \\     
A549, HeLa, MDA            &   Gold  &  Transferrin   & 15,30,{\bf 45}  &  \citenum{Albanese2011}  \\   
Caco-2                     &   Liposomes & Not specified   &  {\bf 40},72,86,97,162    &  \citenum{Andar2013}  \\
\hline
\end{tabular}
\end{center}
\end{table}

In this paper we make a step in this direction. Our main purpose is to
demonstrate that, contrary to the current understanding, the
size-dependence of cellular uptake of NPs is determined by the coat
assembly. To do so, we use a previously developed coarse-grained model
of the coat assembly that focuses on vesicle formation during clathrin-mediated endocytosis
\cite{Banerjee2012a}.  The model was developed to explain the fates
and lifetimes of clathrin coated pits (CCPs); here we use it to study
the size dependence of NP uptake. In particular, we calculate the
dependence of the NP internalization probability and the
internalization time on the NP size. Our results show that the
above-mentioned experimental observations, namely, (1) the optimal NP
size, (2) an upper bound on the size of NPs that can be internalized, 
and (3) the weak dependence of these sizes on type of cells,
NPs, and ligands used in different experiments, can be understood to
be the consequence of the dynamics of coat assembly. This is the main 
result of our work.

\begin{figure}[b]
\centering{\includegraphics[width=4.2in]{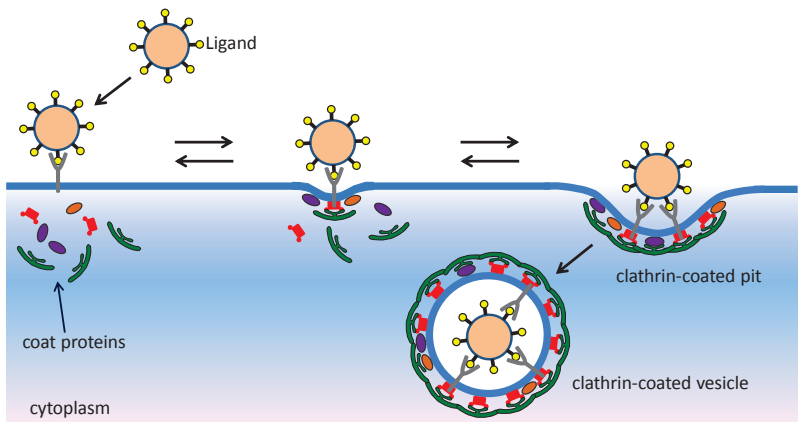}}
\caption{Schematic diagram of NP internalization via clathrin-mediated
  endocytosis. A NP first binds to a specific cell surface receptor forming
a NP-receptor complex. The complex binds the coat
  proteins and CCP assembly begins. The CCP either grows to form a
  vesicle, in which case the NP is internalized, or grows only up to a
  certain size and then disassembles. }
\label{fig1}
\end{figure}

\section*{MODEL}
We start by describing the main steps involved in NP internalization
via clathrin-mediated endocytosis.  During clathrin-mediated endocytosis a ligand-coated NP first binds to a specific
receptor on the surface of the cell membrane (see Fig.\,\ref{fig1}).
After binding, the NP-receptor complex binds adaptor proteins
(typically AP-2), which recruit other endocytic proteins, and
clathrin-coated pit (CCP) assembly begins. CCP assembly is a
stochastic process which has two possible final outcomes
\cite{Ehrlich2004,Loerke2009}. One is that the CCP grows in size and
forms a vesicle, in which case the NP is completely wrapped and
internalized.  The other possibility is that the CCP grows only up to
a certain size and then disassembles.  In this case the NP-receptor
complex becomes free of coat proteins, and the assembly process starts
once again.  We assume that the binding of a NP to the cell membrane
is irreversible, and that the dissociation of the NP-receptor complex
may be neglected. We discuss the restrictions of this assumption
below.

\subsection*{{\bf Quantifying internalization efficiency}}

Similar to other approaches \cite{Gao2005,Decuzzi2007}, we
characterize the NP internalization efficiency by the mean
internalization time, $\tau$, defined as the mean time between the
binding of a NP to the membrane and its internalization. As shown in
Appendix A,
\begin{equation}
\tau =  \tau_w + 
(\tau_0 + P_f\tau_f)/P_w,
\label{eq:tau1}
\end{equation}
where $\tau_0$ is the mean time required for the initiation of CCP
assembly around a free NP-receptor complex, $P_w$ and $P_f = 1- P_w$
denote the probabilities of successful and unsuccessful wrapping of a
NP, and $\tau_w$ and $\tau_f$ denote the mean durations of these
processes.  Here $w$ and $f$ indicate successful and failed wrapping
of the NP, respectively.  Our assumption that the NP-receptor
dissociation may be neglected is valid if $P_w$ is not too
small. Otherwise, $\tau$ becomes very large, and the dissociation of
the complex should be taken into consideration.

\begin{figure}[h]
\centering{\includegraphics[width=3.5in]{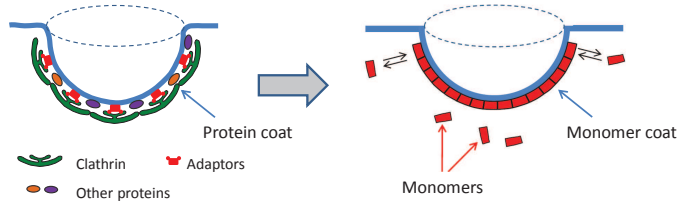}}
\caption{Coarse-grained model of a clathrin-coated pit (CCP). In a
  real CCP the protein coat contains clathrin and several other
  proteins. The model coat is made of identical monomeric units. The
  shape of the model CCP is assumed to be a spherical cap, and the
  average area of a monomer is chosen to be the same as that occupied
  by a clathrin triskelion in a real CCP.}
\label{fig:model}
\end{figure}

In order to calculate the quantities appearing in Eq.\,\ref{eq:tau1}
we use the model of CCP assembly developed in
Ref.\,\citenum{Banerjee2012a}.  Here we briefly describe the model and list
the underlying assumptions.  As mentioned earlier the coat that forms
during CCP assembly contains several proteins and has a complex
structure. Proteins like epsin and amphyphysin bind directly to the
cell membrane and impart a local curvature; whereas clathrin
triskelions (three-legged, pinwheel-wheel shaped complexes) bind with
other clathrin triskelions to form a three-dimensional scaffold which
is linked to the membrane through the adaptor proteins (typically
AP-2). The clathrin scaffold imparts global curvature to the cell
membrane.  Incorporating this complex structure of the coat into a
model is an extremely complicated task. To overcome this difficulty,
in Ref.\,\citenum{Banerjee2012a} we proposed a coarse-grained description
of CCP assembly.  The main idea was to replace the real protein coat
by a coat made up of identical units referred to as monomers
(Fig.\,\ref{fig:model}). We assumed that (1) the coat made up of
monomers has its own bending rigidity and a spontaneous curvature, (2)
the shape of the model CCP (pit) is a spherical cap, (3) the monomers
are structureless, which means that at the time of binding the
orientation of a monomer is not important. Due to these assumptions
certain details of CCP assembly were lost, but this is the price we
had to pay for a tractable model which still contained the essential
features of the assembly process. We validated this model by showing
that it was capable of explaining the experimentally measured lifetime
distribution of CCPs \cite{Banerjee2012a}. Similar coarse-grained
approaches for modeling the protein coat have been used in 
Refs.\,\citenum{Liu2009} and \citenum{Foret2008} for studying 
endocytic vesicle formation in yeast and COP vesicle formation in 
the Golgi, respectively. 

\begin{figure}[b]
\centering{\includegraphics[width=3.5in]{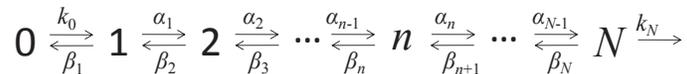}}
\caption{Kinetic scheme of pit assembly. Symbols $n$, and $N$ refer to
  the number of monomers in a pit and a vesicle, respectively.  The
  rate constants $\alpha_n$ and $\beta_n$ characterize the growth and
  decay rates of a pit of size $n$. $k_0$ is the rate at which the
  first monomer binds to the NP-receptor complex, and $k_N$ is the
  rate of scission of a vesicle from the membrane.}
\label{fig:ks}
\end{figure}

Using the coarse-grained model of CCP discussed above, the dynamics of
CCP assembly around a NP-receptor complex can be described by the
kinetic scheme shown in Fig.\,\ref{fig:ks}. In this kinetic scheme the
symbol $n$ is the number of monomers in a pit that forms around the
NP-receptor complex, and $N$ is the number of monomers needed for a
complete vesicle. $N$ is related to the NP diameter $d_{\!_{NP}}$ by
the relationship
\begin{equation}
N = \pi(d_{\!_{NP}}+ 2l_{b})^2/\lambda, 
\end{equation}
where $l_b$ is the typical length of the receptor-ligand bond between
the NP and the cell membrane, and $\lambda$ is the average area
occupied by a monomer.  The rate constants $\alpha_n$ and $\beta_n$
characterize the growth and decay rates of a pit of size $n$,
$k_0=1/\tau_0$ is the rate constant for binding of the first monomer
to the NP-receptor complex, and $k_N$ is the rate constant for the
scission of a vesicle from the membrane.  We assume that the forward
and backward rate constants are related through detailed balance
\begin{equation}
\beta_{n} = \alpha_{n-1} \exp[\tilde{F}(n)-\tilde{F}(n-1)] ,
\quad n = 2,3...,N,
\label{eq:db}
\end{equation}
where $\tilde{F}(n) = F(n)/(k_{\!_{B}} T)$, $F(n)$ is the formation
free energy for a pit containing $n$ monomeric units, $k_{\!_{B}}$ is
the Boltzmann constant, and $T$ is the absolute temperature.

We choose the forward rate constants to be of the form $\alpha_n =
\gamma f(n,N)$, $n= 1,2,...,N-1$, where $\gamma$ is a kinetic
parameter proportional to the product of the free monomers
concentration and the bimolecular association rate constant between a
free monomer and a pit. The function $f(n,N)$ gives the number of
available binding sites on the edge of a pit of size $n$.  Using that
the shape of a pit is a spherical cap, this function can be written as
\begin{equation}
f(n,N) = \rho \sqrt{n(N-n)/N},
\end{equation}
where $\rho$ is a dimensionless parameter  (see Appendix B).

In terms of the kinetic scheme in Fig.\,\ref{fig:ks}, the quantities
$P_w$ and $P_f$ (see Eq.\,\ref{eq:tau1}) are the probabilities that a
random walk, starting from site $n=1$, eventually reaches sites $n=N$
and $n=0$, respectively, and the times $\tau_w$ and $\tau_f$ are the
mean durations of the two processes (which formally are conditional
mean first-passage times \cite{VanKampen2011,Redner2001}).  Analytical
expressions for these quantities are well known \cite{VanKampen2011}
\begin{equation}
P_w = \frac{\Psi_1}{\Psi_1+\sum\limits_{m=1}^{N}\Phi_m}, \quad 
P_f = 1 - P_w,
\label{eq:pw}
\end{equation}

\begin{equation}
\tau_w =
\frac{\sum\limits_{m=1}^{N}\exp[-\tilde{F}(m)]\sum\limits_{l=1}^m\Psi_l
  \sum\limits_{l=m}^{N}\Phi_l}{\Psi_1+\sum\limits_{m=1}^{N}\Phi_m},
\label{eq:tw}
\end{equation}

\begin{equation}
 \tau_f = \frac{\Psi_1\sum\limits_{m=1}^{N}\exp[-\tilde{F}(m)]
   (\sum\limits_{l=m}^{N}\Phi_l)^2}{\sum\limits_{l=1}^{N}\Phi_l(\Psi_1+\sum\limits_{m=1}^{N}\Phi_m)},
\label{eq:tf}
\end{equation}
where functions $\Psi_n$ and $\Phi_n$ are given by $\Psi_n =
\exp[\tilde{F}(n)]/\beta_n$ and $\Phi_n = \exp[\tilde{F}(n)]/\alpha_n$.

\subsection*{{\bf Free energy of pit formation}}
The free energy of pit formation can be written as
\cite{Foret2008,Banerjee2012a} (see details in Appendix C)
\begin{equation}
F(n,N)  = E(N)n  + \sigma\sqrt{n(N-n)/N}.
\label{eq:fe}
\end{equation}
The free energy is mainly dominated by the first term, $E(N)n$, which is 
proportional to the number of monomers in the pit. 
It includes the costs of the membrane and protein coat distortions, 
entropic cost of immobilizing the monomers, and the
binding energy gained due to coat formation. The second term is the
line tension energy with $\sigma$ being the edge-energy constant.  We
use a Helfrich type expression \cite{Helfrich1973} for the membrane
and coat distortion energy, and assume that the spontaneous curvature
of the cell membrane is zero and that of the coat is finite. In
addition, we assume that the binding energy and the entropic cost of
immobilizing the monomers are proportional to the number of monomers
in the pit. Based on these assumptions we get
\begin{equation}
E(N) = \frac{8\pi\kappa_m}{N} + \frac{8\pi\kappa_p}{N}
  \left(1-\sqrt{\frac{N}{N_p}}\right)^2 -\epsilon_b\,,
\label{eq:D}
\end{equation}
where $\kappa_m$ and $\kappa_p$ are the bending
rigidities of the cell membrane and the coat, respectively, $N_p$ is
the natural number of monomers in the coat - which is determined by
the intrinsic coat curvature, and $\epsilon_b$ is the effective
monomer binding energy, i.e., the difference between the binding
energy and entropic cost.

\subsection*{{\bf Parameter values}}

\begin{table}[t]
\caption{Parameter values}
\label{tab2}
\center
\begin{tabular}{lll}
\hline
Parameter      &                   Description                  &    Value    \\ \hline
$\kappa_m$     &      Membrane bending rigidity                 &   $20\,k_{\!_{B}}T$    \\  
$\kappa_p$     &     Protein coat bending rigidity              &   $200\,k_{\!_{B}}T$   \\   
$N_p$          &     Number of monomers in a typical vesicle    &    80         \\   
$\epsilon_b$   &             Binding energy per monomer         &    $10\,k_{\!_{B}}T$   \\ 
$\sigma$       &             Edge energy constant               &    $2\,k_{\!_{B}}T$     \\   
$\gamma$       &                 Kinetic parameter              &     0.18\,sec$^{-1}$  \\   
$\tau_0$       &         Average time for initiation of a CCP          &     20\,sec     \\  
$\beta_1$      &             Backward rate constant             &      0.1\,sec$^{-1}$    \\   
$k_{\!_{N}}$     &                   Vesicle scission rate        &      $\infty$     \\   
$\lambda$      &     Average area occupied by a monomer         &      310\,nm$^2$     \\  
$l_b$      &     Length of receptor ligand bond                 &      15\,nm     \\  
$\rho$      &    Dimensionless parameter      &    2    \\   \hline
\end{tabular}
\end{table} 

The values of the parameters used in our calculation are summarized in
Table\,\ref{tab2}. These values, except for $\epsilon_b$, are
identical to those in Ref.\,\citenum{Banerjee2012a}.  The rationale
behind the choices is as follows: The value of $\kappa_m$ typically
lies between 10-25\,$k_{\!_{B}}T$ \cite{boal2002mechanics}; we choose
$\kappa_m=20$\,$k_{\!_{B}}T$. In vitro, clathrin triskelions assemble
into baskets of different sizes. The size distribution of baskets is
typically narrow and has a peak close to $d_p = 90$\,nm in diameter
\cite{Zaremba1983}.  Using the average area occupied by a clathrin
molecule, $\lambda =310$\,nm$^2$, and the relation $\lambda N_p = \pi
d_p^2$, we find that a 90\,nm basket would have approximately 80
clathrin triskelions. So we choose the natural coat size to be $N_p =
80$.  The value of $\lambda = 310$\,nm$^2$ was estimated using the
relation between diameters of clathrin baskets of different sizes and
the number of clathrin triskelions they contain \cite{Nossal2001}. In
Ref.\,\citenum{Banerjee2012a}, using experimental data on lifetime
distribution of abortive CCPs (CCP with no cargo) we estimated the
value of the effective binding energy per monomer, $\epsilon_b$, to be
approximately 5\,$k_{\!_{B}}T$.  Experiments show that in
the presence of cargo (present case) the binding energy increases.  An
approximate range for its value can be determined using the following
argument: for a clathrin-coated vesicle to be energetically stable,
the effective binding energy has to be greater than the membrane
bending energy ($8\pi \kappa_m \approx 500\,k_{\!_{B}}T$). Since a
typical vesicle has 80 monomers, the binding energy per monomer should
be greater than $500/80 \approx 6\,k_{\!_{B}}T$.  To get an upper
bound to the binding energy we consider the electrostatic binding
energy between proteins containing a BAR (Bin-Amphiphysin-Rvs) domain
and the membranes, which is estimated to be around $15\,k_{\!_{B}}T$
\cite{Parthasarathy2006}. We choose a number in between these values
and set $\epsilon_b = 10\,k_{\!_{B}}T$. The parameter, $\kappa_p$,
captures the effective bending rigidity of the protein coat. 
In Ref.\,\citenum{Banerjee2012a} we
estimated its value to be approximately $\kappa_p = 200\,k_{\!_{B}}T$;
here we use the same value.  To the best of our knowledge, the value
of $\tau_0$ in the case of nanoparticles has never been measured.  In
the case of a particular virus (canine parvovirus) entering via
clathrin-mediated endocytosis, it was found to be approximately
20\,sec \cite{Cureton2012}. Thus we choose $\tau_0 = 20$\,sec. We
choose the length of a receptor ligand bond to be $\l_b=15$\,nm
\cite{Bell1984}.  Parameters $\sigma$, $\rho$, and $\gamma$
have the same values as in Ref.\,\citenum{Banerjee2012a}: 
$\sigma =2\,k_{\!_{B}}T$, $\rho=2$,
$\gamma = 0.18\,\mbox{sec}^{-1}$.

\begin{figure}[b]
\center
\includegraphics[width=3.4in]{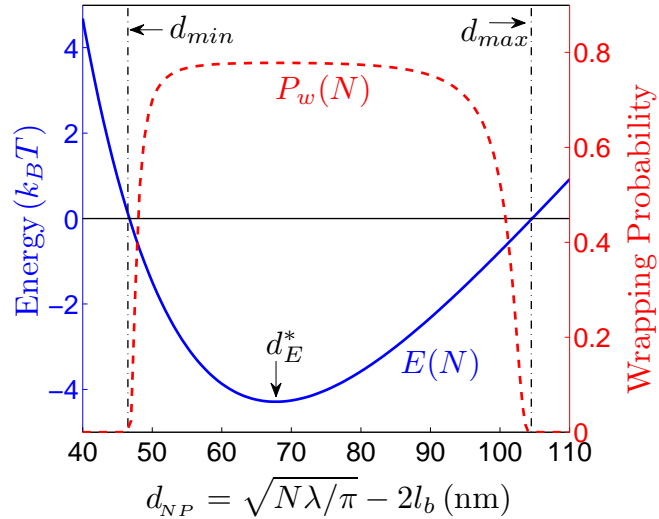}
\caption{Energy $E(N)$, Eq.\,\ref{eq:D}, (solid curve) and the
  wrapping probability, $P_w(N)$, (dashed curve), as functions of the
  NP size. $P_w(N)$ is high when $E(N)<0$, and pit assembly is
  energetically favorable. Dashed-dotted vertical lines at
  $d_{min}\approx 46$\,nm and $d_{max}\approx 105$\,nm correspond to
  the NP sizes at which $E(N)=0$. The arrow at $d_{\!_{E}}^* \approx
  68$\,nm indicates the NP size whose carrier vesicle is energetically
  most stable.}
\label{fig:up}
\end{figure}

\section*{RESULTS}
Figure\,\ref{fig:up} shows $E(N)$, Eq.\,\ref{eq:D}, as a function of
the NP size. It attains a minimum at $d_{\!_{E}}^*\approx 68$\,nm,
which corresponds to the size of the NP whose carrier vesicle is
energetically most stable. This size can be found by solving the
equation $\partial E(N)/\partial N = 0$, which leads to
\begin{equation}
d_{\!_{E}}^*= (1 + \kappa_m/\kappa_p)\sqrt{\lambda N_p/\pi} 
- 2l_b.
\label{eq:ddm}
\end{equation}
For NPs larger than $d_{\!_{E}}^*$, $E(N)$ increases due to the
energetic cost of protein coat deformation, whereas for NPs smaller
than $d_{\!_{E}}^*$, $E(N)$ increases mainly due to the energetic cost
of cell membrane deformation.  The vertical dash-dotted lines at
$d_{min}\approx 46$\,nm and $d_{max}\approx 105$\,nm show the NP sizes
at which $E(N) = 0$. Analytical expressions of these sizes can be
obtained by solving the equation $E(N) = 0$, which leads to
\begin{equation}
\frac{d_{min}}{d_p} = \frac{1 -
  \sqrt{(1+k_m')\epsilon_b'-k_m'}}{(1-\epsilon_b')}- \frac{2l_b}{d_p}, \quad
\frac{d_{max}}{d_p} = \frac{1 +
  \sqrt{(1+k_m')\epsilon_b'-k_m'}}{(1-\epsilon_b')}- \frac{2l_b}{d_p},
\label{eq:ds}
\end{equation}
where $\kappa_m' = \kappa_m/\kappa_p$, $ \epsilon_b' =
N_p\epsilon_b/8\pi\kappa_p$, and $d_p = \sqrt{\lambda N_p/\pi}$. 

Figure\,\ref{fig:up} also shows the plot of the wrapping probability,
$P_w(N)$.  In regions where the sum of membrane and coat distortion
energies is greater than the binding energy ($E(N)>0$), pit assembly
is energetically unfavorable, and $P_w(N)$ is negligibly small.  In
contrast, in the region where $E(N)<0$, $P_w(N)$ rises sharply and
then remains high (about 0.8) and approximately constant. Notably,
sizes of several viruses which enter through clathrin-mediated endocytosis, including dengue
virus (40-60\,nm), semliki forest virus (50-70\,nm), and reovirus
(60-80\,nm), fall within this range \cite{Mudhakir2009}.

\begin{figure}[b]
\center
\includegraphics[width=3.5in]{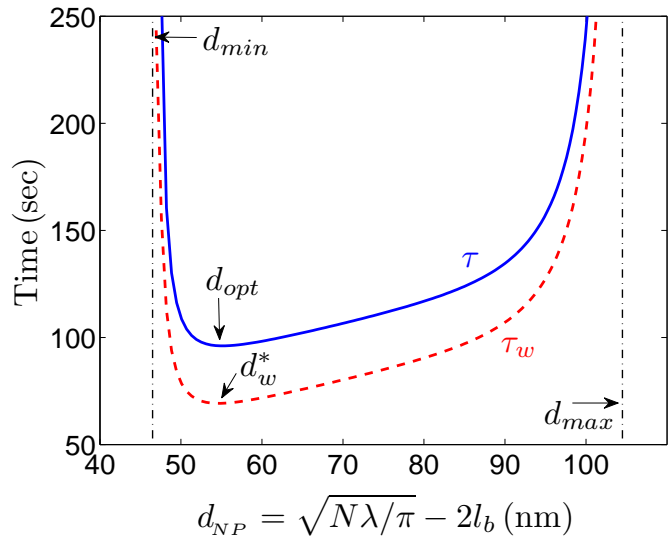}
\caption{Mean wrapping time, $\tau_w$, Eq.\,\ref{eq:tw}, (dashed
  curve), and the mean internalization time, $\tau$, calculated using
  Eq.\,\ref{eq:tau1} with $\tau_0 = 20$\,sec, as functions of the NP
  size. Arrows at $d_{w}^*$ and $d_{opt}$ indicate the NP sizes at
  which $\tau_w$ and $\tau$ are minimum, respectively.  Our estimate
  $d_{opt}\approx 55$\,nm matches well with experimental observations
  presented in Table\,\ref{tab1}. }
\label{fig:mmt}
\end{figure}

Figure\,\ref{fig:mmt} shows plots of the mean wrapping time, $\tau_w$,
Eq.\,\ref{eq:tw}, and the mean internalization time, $\tau$,
Eq.\,\ref{eq:tau1}, as functions of the NP size. The mean wrapping
time has a minimum at $d_{w}^*\approx 55$\,nm, which is different from
$d_{\!_{E}}^*\approx 68$\,nm. The difference between the two sizes is
due to the NP size-dependent dynamics of coat assembly. At $d_{w}^*$, 
$E(N)\approx
-3\,k_{\!_{B}}T$, and hence $\alpha_n/\beta_n \approx 20\gg 1$. This
implies that coat assembly (hence NP wrapping) proceeds with a low 
probability of monomer dissociation. As a consequence, the wrapping time 
is determined by the rate of arrival of a new monomer. For NPs larger 
than $d_{w}^*$, up to
approximately 85\,nm, the inequality $\alpha_n/\beta_n \gg 1$ holds
true, but the number of monomers needed to wrap a NP increases. As a
consequence, the wrapping time increases with NP size. For NPs larger
than 85\,nm, the dissociation of monomers becomes more frequent, which
causes the wrapping time to increase drastically.  For NPs smaller
than $d_{w}^*$, the free energy $E(N)$ rises sharply, and, therefore,
the rate constants $\alpha_n$ and $\beta_n$ become comparable. Thus,
even though the number of monomers needed to wrap a NP decreases, the
wrapping time increases due to frequent dissociation of the monomers.
The mean time of failed attempts, $\tau_f$, given in Eq.\,\ref{eq:tf},
shows a trend very similar to that of $\tau_w$, but its values are
almost an order of magnitude smaller.

In Fig.\,\ref{fig:mmt} we show the mean internalization time, $\tau$,
calculated using Eq.\,\ref{eq:tau1} with $\tau_0 = 20$\,sec.  The mean
internalization time includes the mean wrapping time, $\tau_w$, and
the mean time spent by the NP in failed attempts.  In the region where
$P_w$ is high and almost constant (see Fig.\,\ref{fig:up}), we find
$P_w\tau_w \gg P_f\tau_f$. Equation\,\ref{eq:tau1} then simplifies
to $\tau \approx (\tau_0/P_w) + \tau_w$, and $\tau$ is just the mean
wrapping time, $\tau_w$, shifted by a constant $\tau_0/P_w$.  When
$d_{\!_{NP}}$ is close to $d_{min}$ or $d_{max}$, $P_w$ drops sharply
and, therefore, $\tau$ increases more rapidly than $\tau_w$.  The mean
internalization time has a minimum at $d_{opt}$, which corresponds to
the NP size at which the cellular uptake of the NPs is the fastest.
Our analysis predicts $d_{opt}\approx 55$\,nm, which is close to the
optimal NP size observed in different experiments (see Table\,1). In
the case of NPs, the size dependence of internalization times at a
single NP level has never been measured, therefore a direct comparison
of our results with with experimental data is not possible. However,
internalization times of some viruses and virus-like particles
entering into cells via clathrin-mediated endocytosis have been measured to be in the range
50-400\,sec \cite{Ehrlich2004,Rust2004,Cureton2012,Bruin2007}, which 
agrees with our analysis.

\begin{figure}[]
\center
\includegraphics[width=3.1in]{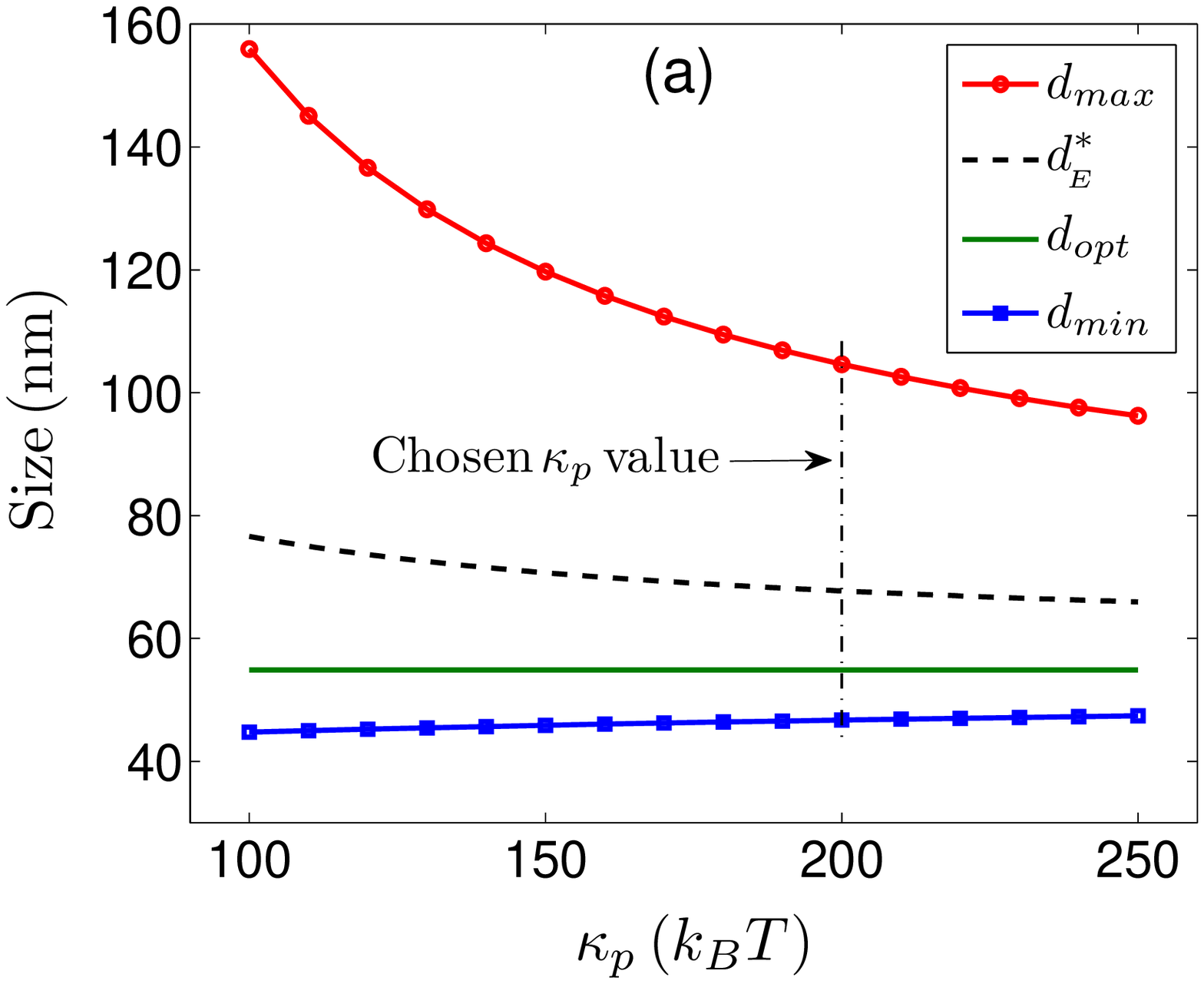}
\vspace{0.25in}

\includegraphics[width=3.1in]{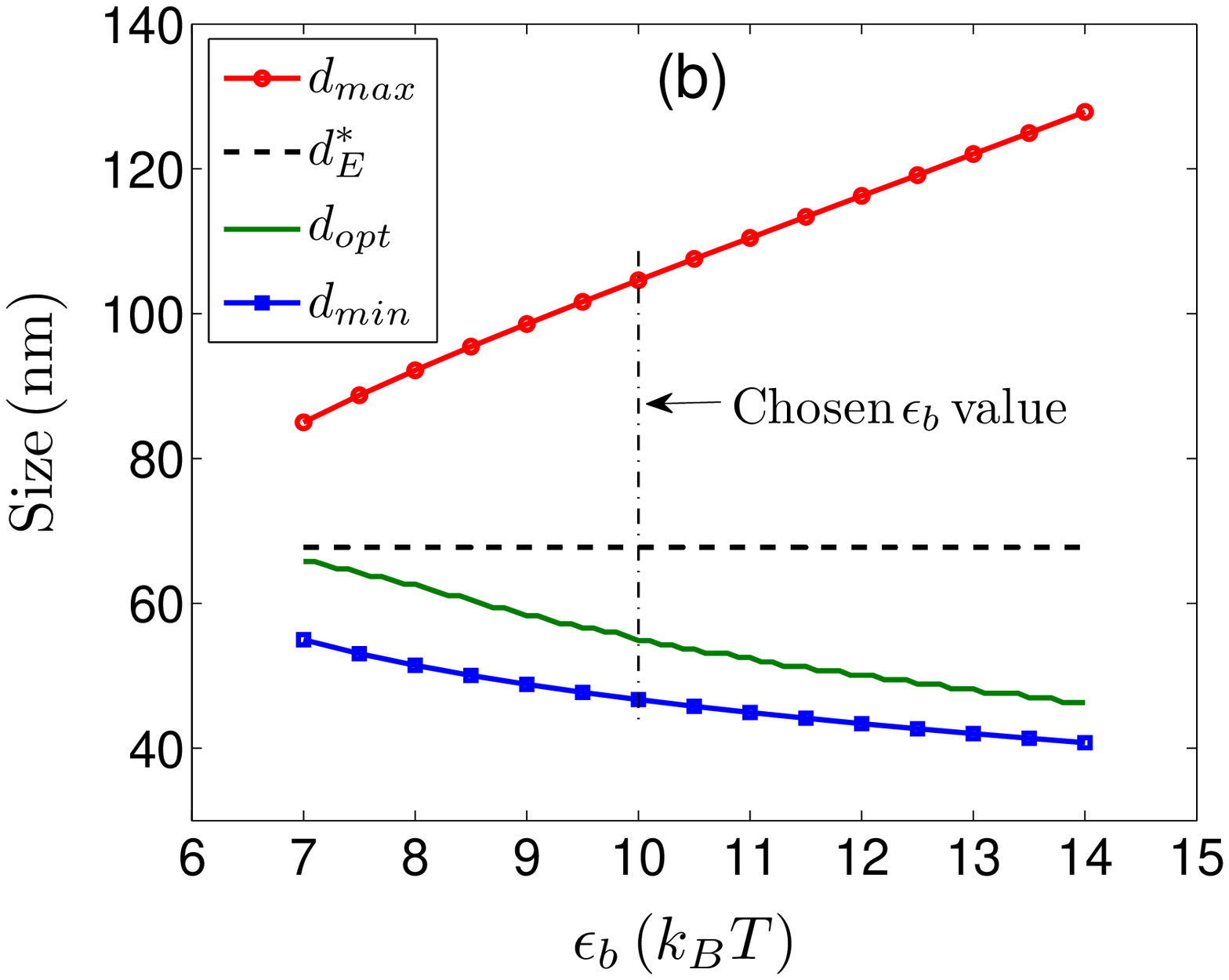}
\vspace{0.25in}

\includegraphics[width=3.1in]{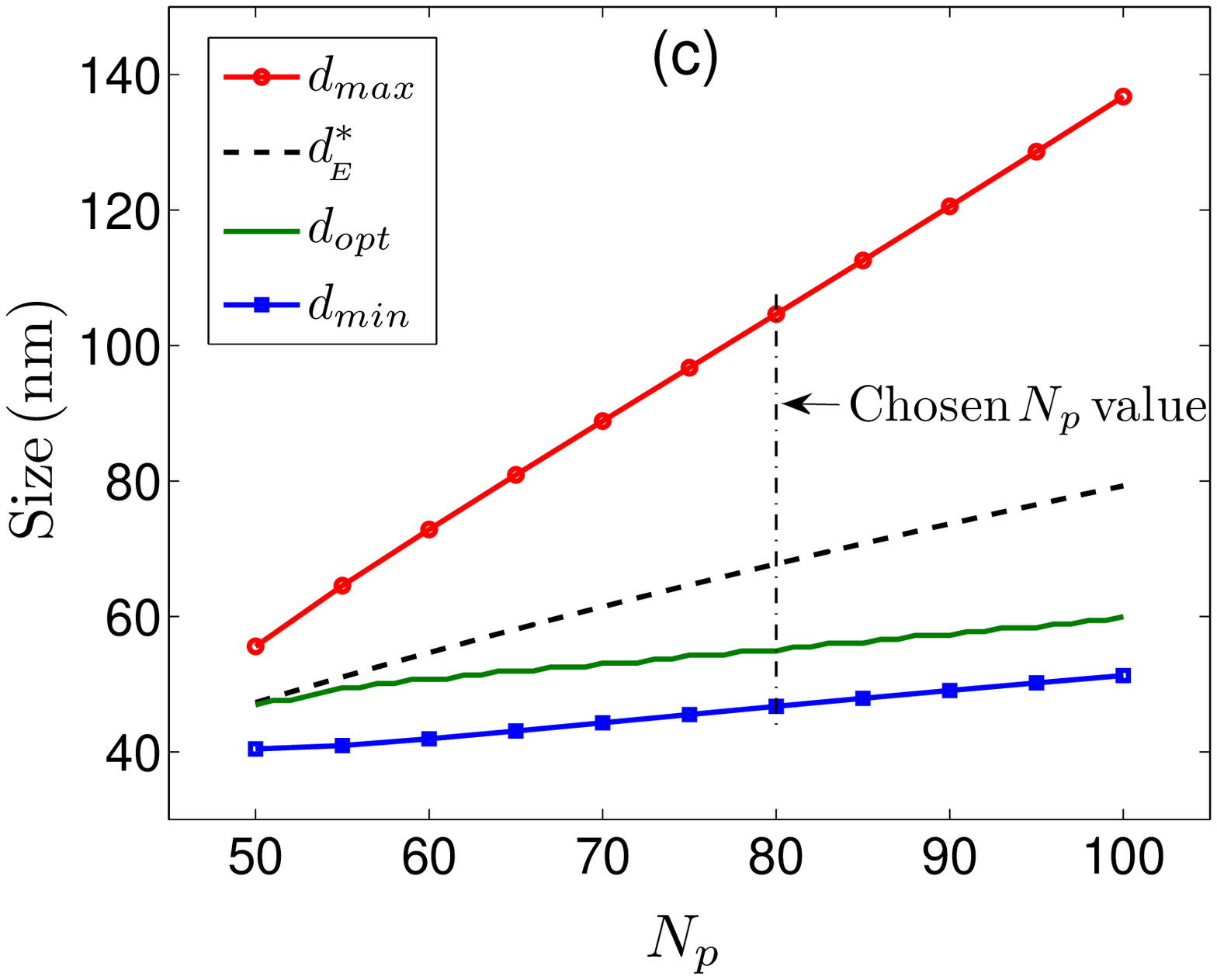}
\caption{Plots of $d_{min}$, $d_{max}$ (Eq.\,\ref{eq:ds}),
  $d_{\!_{E}}^*$ (Eq.\,\ref{eq:ddm}), and $d_{opt}$ (calculated
  numerically), as functions of (a) $\kappa_p$ - the bending rigidity
  of the protein coat (b) $\epsilon_b$ - effective monomer binding energy, and
  (c) $N_p$ - natural number of monomers in the coat. The vertical
  dash-dotted line in each case shows the parameter value used in our
  calculations. The plot shows that our results for $d_{min}$,
  $d_{\!_{E}}^*$, and $d_{opt}$ are stable with respect to small
  variations in the parameter values. Only the maximum NP size
  $d_{max}$ shows significant variation. }
\label{fig:DvK}
\end{figure}

Finally, we look at sensitivity of our results to variations in coat
parameter values. In Fig.\,\ref{fig:DvK} we show plots of $d_{min}$,
$d_{max}$, $d_{opt}$, and $d_{\!_{E}}^*$ as functions of the
parameters $\kappa_p$, $\epsilon_b$, and $N_p$. The dependence of the
optimal size, $d_{opt}$, on the coat parameters has been determined
numerically, while other dependences are are given by Eqs.\,\ref{eq:ddm} 
and \ref{eq:ds}.  Only the maximum NP size, $d_{max}$, shows appreciable
variation; the other quantities show weak dependences on the coat
parameters.  This demonstrates that our main results are stable with
respect to small variations in the coat parameters around their chosen
values.

\section*{SUMMARY AND DISCUSSION}

In this study we investigated how the assembly of the protein coat on
the cytoplasmic side of the plasma membrane affects the cellular
uptake of NPs. To address this question we have used a previously
developed model of clathrin-coated vesicle formation. We have used the
mean internalization time of a NP, $\tau$, as the measure of its
internalization efficiency, and calculated the dependence of $\tau$ on
the NP size. We found that the NP size has lower and upper boundaries
($d_{min}$ and $d_{max}$) at which the internalization time becomes
very large, i.e., beyond these sizes, internalization via clathrin-mediated 
endocytosis is highly improbable. 
We also found that there is an optimal NP size $d_{opt}$ at
which the internalization time is a minimum. All these sizes are
determined by the parameters of the coat assembly process.

As described earlier,
$d_{opt}$ is determined by the dynamics of coat assembly. Since the 
coat parameters do not change appreciably between different cells and 
are also independent of the details of the NP design, an 
explanation for why the same optimal size was observed in different
experiments (see Table\,\ref{tab1}) follows 
naturally from our analysis. In contrast, this observation is
difficult to rationalize using previous models which predict that
optimal size depends on the ligand density on the NP, the density of the
corresponding receptors on the cell membrane, and the receptor-ligand
binding energy \cite{Gao2005,Zhang2009,Yuan2010}, since these
parameters can vary significantly depending on the cell line and
ligand used in the experiment.

Our calculation of the smallest NP size, $d_{min}\approx 46$\,nm, is
based on the assumption that the size of the NP is related to the size
of its carrier vesicle through $d_{V}^{min} = d_{min} + 2l_b$, which
gives $d_{V}^{min}\approx 76$\,nm. In principle, however, NPs with
diameter slightly smaller than $d_{min}$ can be internalized in a
vesicle of size $d_{V}^{min}$.  Also, it has been shown that NPs much
smaller than $d_{min}$ can be internalized in clusters (multiple
particles per vesicle), in a vesicle of size much larger than the size
of individual NPs \cite{Chithrani2007}.  Therefore, a comparison of
$d_{min}$ with experimental data is meaningless. In this case it is
more meaningful to compare $d_{V}^{min}$ with the size of the smallest
clathrin-coated vesicles.  Experimentally observed size of smallest 
clathrin-coated vesicles is approximately
70\,nm in diameter \cite{Kirchhausen2009}, which agrees very well with
our estimate.
 
Our estimate of the largest NP size, $d_{max}\approx 105$\,nm, and the
size of its carrier vesicle, $d_{V}^{max}\approx 105+30= 135$\,nm, are
smaller than their corresponding experimentally measured values
200\,nm \cite{Rejman2004d,Oh2009}, and 200\,nm \cite{Kirchhausen2009},
respectively. As shown in Fig.\,\ref{fig:DvK} the value of $d_{max}$
is sensitive to the coat parameter values. By changing their values
slightly we can get $d_{max}$ close to experimentally observed values,
while keeping $d_{opt}$ the same.  For example, for $\kappa_p = 150
k_{\!_{B}}T$ and $\epsilon_b = 12 k_{\!_{B}}T$, we get $d_{max} =
140$\,nm, $d_V^{max} = 170$\,nm, $d_{min} = 40$\,nm, and
$d_{opt}=50$\,nm. However, in this paper our aim is not to match the
different sizes precisely, but rather to see the extent to which our
previously developed model can explain the size dependence of NP
uptake without changing the parameter values. Considering the fact
that most of the parameters were determined in a completely different
context (by fitting lifetime distribution of abortive CCPs) we think
that such a disparity is acceptable. As mentioned
earlier, other models that do not take coat assembly into
consideration, incorrectly predict that very large (micron size) NPs
can be internalized via receptor-mediated endocytosis \cite{Gao2005,Zhang2009}.

Although in our model the density of ligands on a NP, the receptor 
density on the cell membrane, and the receptor-ligand binding energy 
do not appear explicitly, these factors do enter our model implicitly. 
For example, our initial assumption 
that the dissociation of the NP from the membrane can be neglected 
would hold true only if either the receptor-ligand binding energy is strong, 
or the ligand and receptor densities are large enough so that a NP 
quickly attaches to the cell membrane by multiple receptor-ligand bonds. 
Multiple bonds lead to an increased lifetime of  NPs on the cell 
membrane \cite{Hong2007,Jiang2008}. Also, it has been shown that 
CCP assembly is triggered when there is receptor 
clustering \cite{Liu2010} which, for our model, implies that a few 
receptor-ligand bonds probably have to form before the first monomer 
can arrive. Thus, the time $\tau_0$ might be affected by the above 
mentioned parameters.

Experimentally, the question of how the ligand density on a NP and 
receptor density on cell membrane affect cellular
uptake is not clearly understood. It has been observed that increasing
the ligand density increases cellular uptake due to an enhanced
residence time of the NP on the cell membrane, and not due to an
increase in the internalization rate \cite{Hong2007}. This observation
is consistent with our hypothesis that the internalization time is
determined by the kinetics of coat assembly.  The overall
picture of NP internalization proposed in our model can be tested
experimentally with total internal reflection fluorescence (TIRF)
microscopy.  Using dual color TIRF, where both NPs and CCPs are
fluorescent, uptake of NPs at a single particle level can be
monitored. This will allow simultaneous measurements of $\tau_0$,
$\tau_w$, and $\tau$.

To conclude we show that several experimental observations related to
size dependent cellular uptake of NPs, including the optimal NP size,
can be understood to be consequences of the protein coat assembly
process.  Therefore, future efforts on modeling endocytosis of NPs and
designing NPs for biomedical applications must take the effect of the
protein coat assembly explicitly into consideration.
  
\section*{APPENDICIES}

\subsection*{{\bf Appendix A: Mean internalization time $\tau$}}
Here we derive the expression for the mean internalization time $\tau$
given in Eq.\,\ref{eq:tau1}. This time is defined as
the average time between the NP binding to the cell membrane and its
internalization, assuming that the binding is irreversible.  Consider
an ensemble of NP-receptor complexes formed on the cell membrane at
time $t = 0$. The mean internalization time $\tau$ can be written as
\begin{equation}
\tau  = P_w(\tau_0 + \tau_w) + P_fP_w(2\tau_0 + \tau_f + \tau_w) +
P_f^2P_w(3\tau_0 + 2\tau_f + \tau_w) + .....  
\label{eq:tau}
\end{equation}
The first term in right hand side of this equality is the contribution
from the fraction ($P_w$) of complexes which are internalized on the
first attempt, i.e., the complexes which bind to coat protein and get
internalized. The second term is the contribution from the fraction
($P_fP_w$) of complexes that are internalized on the second attempt,
i.e., they bind to coat proteins, dissociate from them, bind to coat protein for the
second time, and then get internalized. Subsequent terms can be
understood in the same way.  Upon summing the series, we obtain 
\begin{equation}
\tau = (\tau_0 + P_w\tau_w + P_f\tau_f)/P_w\,.
\end{equation}
This expression for the mean internalization time $\tau$ can be written 
in the form given in Eq.\,1.

\subsection*{{\bf Appendix B: Derivation of $f(n)$ in Eq.\,4}}
\begin{figure}[b]
\center
\includegraphics[width=2.5in]{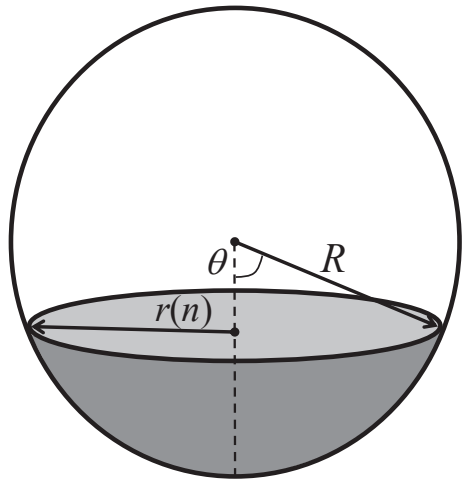}
\caption{Spherical cap model of a pit.}
\label{fig:sc}
\end{figure}

Using spherical coordinates (see Fig.\ref{fig:sc}) the surface area of the
pit can be written as
\begin{equation}
A(\theta) = 2\pi R^2[1-\cos(\theta)] = \lambda n,
\end{equation}
where $R$ is the radius of a sphere having the same curvature as the pit. 
This leads to the relation between $\cos(\theta)$ and the number of
monomers, $n$, in the pit,
\begin{equation}
\cos(\theta)  = 1-\frac{\lambda n}{2\pi R^2}\,.
\end{equation}
From the above equation, we thus infer that the radius, $r(n)$, of the circular
growing edge of a pit is
\begin{equation}
r(n) = R\sin(\theta) = R\sqrt{[1+\cos(\theta)][1-\cos(\theta)]}=
\sqrt{\left(\frac{\lambda n}{\pi}\right)\left(1-\frac{\lambda n}{4\pi
    R^2}\right)}.
\end{equation}
Introducing the average linear span of the monomer, denoted by $L$, we
find that the number of available binding sites on the periphery of
the pit is
\begin{equation}
f(n) = \frac{2\pi r(n)}{L} = \frac{2\pi}{L}\sqrt{\left(\frac{\lambda
    n}{\pi}\right)\left(1-\frac{\lambda n}{4\pi R^2}\right)}.
\label{fn}
\end{equation}
By changing variables from $R$ to $N$ using the relation $4\pi R^2 =
\lambda N$, we arrive at
\begin{equation}
f(n) = \rho \sqrt{n(N-n)/N},
\end{equation}
where $\rho$ is a dimensionless parameter given by $\rho = \sqrt{4\pi
  \lambda/L^2}$.

\subsection*{{\bf Appendix C: Derivation of $F(n)$ in Eq.\,8}}
The formation free energy of a pit made of $n$ monomers and having a 
curvature $c$ can be written as 
\begin{equation}
F(n,c) =  2\kappa_m\lambda nc^2 + 2\kappa_p \lambda
  n(c-c_p)^2 - \epsilon_b n +\sigma f(n,c).
  \label{te1}
\end{equation}
The first term is the Helfrich energy \cite{Helfrich1973} describing
the energetic cost of bending the cell membrane assuming that its
spontaneous curvature is zero. The second term represents the bending
energy of the protein coat, with $c_p$ being the spontaneous curvature
of the coat. The third term represents the effective binding energy.
The fourth term is the line tension energy. By changing variables from
$c$ to $N$ and $c_p$ to $N_p$ using the relations $4\pi /c^2 = \lambda
N$ and $4\pi /c_p^2 = \lambda N_p$\,, we arrive at the expression
for the free energy of the pit formation given in Eq.\,\ref{eq:fe}.

\section*{ACKNOWLEDGMENTS}
This study was supported by the Intramural Research Program 
of the National Institutes of Health (NIH) -- {\it Eunice Kennedy Shriver}
National Institute of Child Health and Human Development, and the
Center for Information Technology.




\end{document}